# Energy Spectra for Fractional Quantum Hall States

Shosuke Sasaki

*Shizuoka Institute of Science and Technology 2200-2 Toyosawa Fukuroi, 437-8555, Japan*
*E-mail: sasaki@ns.sist.ac.jp*

Fractional quantum Hall states (FQHS) with the filling factor ν = *p/q* of *q* < 21 are examined and their energies are calculated. The classical Coulomb energy is evaluated among many electrons; that energy is linearly dependent on 1/ν. The residual binding energies are also evaluated. The electron pair in nearest Landau-orbitals is more affected via Coulomb transition than an electron pair in non-nearest orbitals. Each nearest electron pair can transfer to some empty orbital pair, but it cannot transfer to the other empty orbital pair because of conservation of momentum. Counting the numbers of the allowed and forbidden transitions, the binding energies are evaluated for filling factors of 126 fraction numbers. Gathering the classical Coulomb energy and the pair transition energy, we obtain the spectrum of energy versus ν. This energy spectrum elucidates the precise confinement of Hall resistance at specific fractional filling factors.

1. **Classical Coulomb Energy of FQHS**

   The fractional quantum Hall effect is caused intrinsically by electron-electron interactions. Many theoretical approaches have been carried out to clarify the phenomenon [1–5]. Hitherto, many investigations for FQH states have been undertaken, but there are only a few theories that address the energy spectrum structure, one of which is the theory of Halperin [4], whose results show many cusps in energy versus ν. We calculate the classical Coulomb energy of many electron states in Landau levels, and also evaluate binding energies via residual Coulomb transitions using the improved method of Tao and Thouless [6]. The result shows a discrete energy spectrum versus ν.

   A two-dimensional electron system in a quantum Hall device obeys the Hamiltonian $H_0$, when the Coulomb interactions among electrons are neglected.

$$H_0 = (\mathbf{p} + e\mathbf{A})^2/(2m) + U(y) + W(z), \quad \mathbf{A} = (-yB,\ 0,\ 0)$$

Therein, directions *y* and *z* are respectively denote directions of the Hall voltage and magnetic field; $U(y)$ and $W(z)$ are confining potentials in directions *y* and *z*. Eigenfunctions of $H_0$ are the Landau states $\psi_k$ as

$$H_0\psi_k = E_0\psi_k, \psi_k = \psi_k(x,y,z) = ue^{ikx}e^{-\alpha(y-c)^2}\phi(z), [-\frac{\hbar^2}{2m}\frac{\partial^2}{\partial z^2} + W(z)]\phi(z) = \lambda\phi(z)\ldots(1)$$

$$E_0(k) = \lambda + U(c) + \frac{\hbar eB}{2m} = b + \frac{\hbar eB}{2m}\ldots(2)$$

$$c = \frac{k\hbar}{eB}\ldots(3)$$

for ground states in Landau levels. Therein, $c$ indicates the center position of the electron in the $y$ direction. In addition, $E_0(k)$ is the single electron energy (2) with wave number $k$. In the right hand side of Eq. (2), $b$ is a constant value because the functional form of potential $U$ is uniform for a good quality device except for both ends of the device. The total Hamiltonian of many electrons is expressed as the following.

$$H_T = \sum_{i=1}^{N} H_0(x_i, y_i, z_i) + H_C, \quad H_C = \sum_{i=1}^{N-1} \sum_{j>i}^{N} \frac{e^2}{4\pi\epsilon\sqrt{(x_i - x_j)^2 + (y_i - y_j)^2 + (z_i - z_j)^2}} \ldots (4)$$

The first-order wave function is

$$\Psi(k_1, \cdots, k_N) = \frac{1}{\sqrt{N!}} \begin{vmatrix} \psi_{k_1}(x_1, y_1, z_1) & \cdots & \psi_{k_1}(x_N, y_N, z_N) \\ \vdots & & \vdots \\ \psi_{k_N}(x_1, y_1, z_1) & \cdots & \psi_{k_N}(x_N, y_N, z_N) \end{vmatrix}. \quad (5)$$

The sum of the single electron energies and the classical Coulomb energies is

$$W(k_1, k_2, \ldots, k_N) = \sum_{i=1}^{N} E_0(k_i) + C(k_1, k_2, \ldots, k_N) \ldots (6)$$

$$C(k_1, \ldots, k_N) = \int \cdots \int \Psi(k_1, \ldots, k_N)^* H_C \Psi(k_1, \ldots, k_N) dx_1 dy_1 dz_1 \ldots dx_N dy_N dz_N$$

The total Hamiltonian $H_T$ is divisible into the diagonal component $H_D$ and the off-diagonal component $H_I$ as [7]

$$H_D = \sum_{k_1, \ldots, k_N} |\Psi(k_1, \ldots, k_N)\rangle W(k_1, \ldots, k_N) \langle \Psi(k_1, \ldots, k_N)|, \quad H_I = H_T - H_D \ldots (7)$$

The eigenenergy $W(k_1, \cdots, k_N)$ depends upon the wave numbers $(k_1, \cdots, k_N)$ and takes the minimum value at the most uniform distribution of center positions $(c_1, \cdots, c_N)$.

We examine the state with a filling factor of 3/5 as one example. Figure 1 shows the most uniform distribution at $\nu = 3/5$. Therein, the straight lines are filled with electrons and the dotted lines indicate empty orbitals. Electrons A and B occupy the nearest neighbor orbitals; electrons B and C occupy the second-nearest orbitals. We describe the classical Coulomb energy as $\xi$ for the electron pair AB, and that of the second-nearest pair by $\eta$.

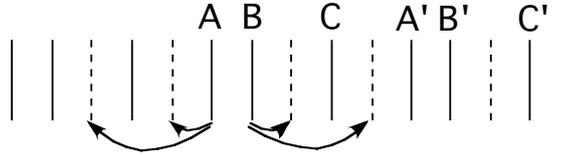

Fig. 1. Most uniform distribution of electrons at $\nu = 3/5$. Dotted lines indicate empty orbitals.

The classical Coulomb energy between electron A and electron C becomes negligibly small because of the screening effect of electron B. Therefore, we obtain the main part of the classical Coulomb energy $C(k_1, \cdots, k_N)$ for $\nu = 3/5$ as the following.

$$C(\nu = 3/5) \approx \xi N/3 + 2\eta N/3$$

Therein, $N$ is the total number of electrons. For a general case of $\nu = p/q$ ($\frac{1}{2} < \nu < 1$), $p$ electrons exist for $q$ orbitals in a unit cell and $q - p$ orbitals are empty in a unit cell. Considering the condition $\frac{1}{2} < \nu < 1$, there are $q - p$ second-nearest electron pairs for each unit cell. These second-nearest pairs have energy $\eta$ per pair. The number of the nearest electron pairs is $2p - q$. These nearest pairs have energy $\xi$ per pair. Consequently, the classical Coulomb energy $C(k_1, \cdots, k_N)$ for $\nu = p/q$ ($\frac{1}{2} < \nu < 1$) is

$$C(\nu = p/q) \approx (2p - q)\xi N/p + (q - p)\eta N/p. \tag{8}$$

The total eigenenergy of $H_D$ is the sum of the single electron energies and the classical Coulomb energy $C(k_1, \cdots, k_N)$ as

$$W(\nu = p/q) \approx (b + \hbar eB/(2m))N + (2\xi - \eta)N - (\xi - \eta)Nq/p.$$

The energy per electron, namely $W/N$ is obtained as

$$W/N \approx \beta + \hbar eB/(2m) - (\xi - \eta)/\nu, \text{ where } \beta = b + (2\xi - \eta). \tag{9}$$

This equation shows that the classical Coulomb energy per electron depends linearly on $1/\nu$.

## 2. Spectrum of FQHS Energy per Electron

The energy via interaction $H_I$ is calculated using the second-order perturbation method. The electron pair occupying the nearest orbitals is most strongly affected by the interaction $H_I$. We count the binding energy of nearest electron pair. A value $Z$ is defined by the following summation:

$$Z = -\sum_{\Delta k \neq 0, -2\pi/l} \frac{\langle k_A, k_B | H_I | k'_A, k'_B \rangle \langle k'_A, k'_B | H_I | k_A, k_B \rangle}{W_G - W_{\text{excite}}(k_A \to k'_A, k_B \to k'_B)} \cdots (10)$$

This summation is carried out for all wave number transfers ($\Delta k = k'_B - k_B$) except 0 and $2\pi/\ell$, and approximates to integration because of the extremely small value of $2\pi/\ell$ for a usual device length.

We explain the perturbation calculation for an example $\nu = 3/5$. In the electron pair AB of Fig. 1, the electron A transfers to the third orbital to the left when the electron B transfers to the third orbital to the right because of the momentum conservation $k'_A + k'_B = k_A + k_B$ and Eq. (3). The binding energy of the second-order perturbation of $H_I$ is $(2/5)Z$ for the electron pair AB because the transitions are allowed to two vacant orbitals and are forbidden to three occupied orbitals for each unit cell. This binding energy is shared with three electrons. Therefore, the binding energy per electron is $\omega_e(\nu = 3/5) = (2/15)Z$. Similarly, we can calculate the binding energies $\omega(\nu)$ for various fractions $\nu$. Then, the total energy per electron $\varepsilon(\nu)$ is obtained by subtracting the binding energy $\omega(\nu)$ from $W/N$ (see Eq. (9)), as follows.

$$\varepsilon(\nu) = \beta + \hbar eB/(2m) - (\xi - \eta)/\nu - \omega(\nu) \tag{11}$$

We draw the schematic behavior of the function $\varepsilon(\nu)$ versus $1/\nu$ in Fig. 2. Therein, we can see many local minima.

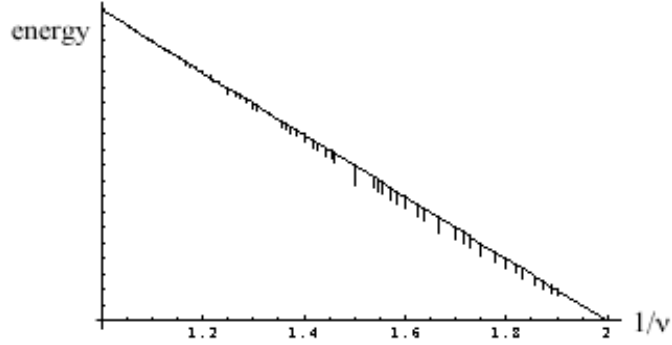

Fig. 2. FQHS energy per electron versus $1/\nu$ for $\nu = p/q$ and $q<21$.

In the Halperin theory [4], the energy spectrum has many cusps, but is continuous for the filling factor $\nu$. On the other hand, our energy spectrum in this paper is discontinuous for changing of $\nu$; the base line decreases linearly for increasing $1/\nu$, as shown in Fig. 2. Moreover, our energy $\varepsilon(\nu)$ increases linearly according to increments of magnetic field strength $B$, as in Eq. (11). These characteristics produce the precise confinement of Hall resistance for several fractional filling factors. We explain this structure for the case of $\nu = 2/3$ as one example. We draw schematic figures of the energy spectrum near $\nu = 2/3$ in Fig. 3, where $\mu$ is the chemical potential of electrons. All states with $\varepsilon(\nu)<\mu$ are filled with electrons at a low temperature because of Fermi distribution, but the states with $\varepsilon(\nu)>\mu$ are empty.

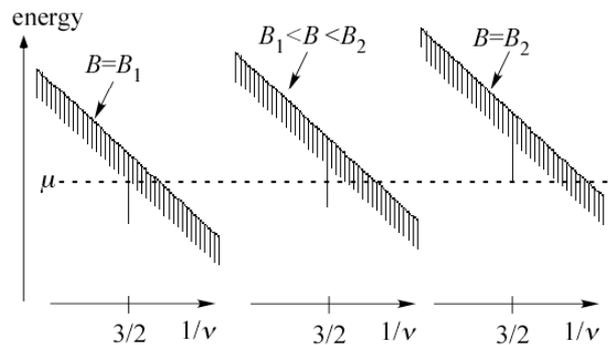

Fig. 3. Schematic figures of energy spectra near $\nu = 2/3$ for three values of $B$

As shown in Fig. 3, energy $\varepsilon(\nu = 2/3)$ is less than $\mu$ for the region of $B_1<B<B_2$. In addition, all states with $\nu > 2/3$ have energy values larger than $\mu$ for $B_1<B<B_2$. Consequently, only the state with $\nu = 2/3$ is filled with electron and the precise confinement occurs for $B_1<B<B_2$. Moreover, the value of $1/\nu$ is approximately proportional to the strength of $B$ because of Eq. (11). These characteristics are in agreement with experimental data.